\documentclass[twocolumn,showpacs,amssymb,10pt]{revtex4}
\usepackage{epsfig}
\usepackage{bm}
\usepackage[latin1]{inputenc}

\def\ketm#1{  \left\vert  #1   \right\rangle   }

\def\mem#1#2#3{  \left\langle #1 \left\vert  #2 \right\vert #3 \right\rangle   }
\def\rmem#1#2#3{  \left\langle #1 \left\vert \left\vert  #2
                  \right\vert \right\vert #3 \right\rangle   }

\def\sixjm#1#2#3#4#5#6{  \left\{ \begin{array}{ccc}
                                               #1 & #2 & #3  \\
                                               #4 & #5 & #6
                     \end{array} \right\}   }

\begin{document}

%
%

\title{Negative continuum effects on the
two--photon decay rates of hydrogen--like ions}

%
%

\author{Andrey Surzhykov$^{1,2}$, José Paulo Santos$^{3,4}$, Pedro Amaro$^{1,3,4}$ and Paul Indelicato$^{5}$}
\affiliation{$^{1}$Physikalisches Institut,
Universität Heidelberg, Philosophenweg 12, D--69120 Heidelberg, Germany\\
$^{2}$GSI Helmholtzzentrum für Schwerionenforschung, Planckstr. 1, D--64291 Darmstadt, Germany \\
$^{3}$Departamento de F\'isica, Faculdade de Ciências e
Tecnologia,
FCT, Universidade Nova de Lisboa, 2829-516 Caparica, Portugal\\
$^{4}$Centro de F\'isica Atómica da Universidade de Lisboa, Av.
Prof. Gama Pinto 2, 1649-003 Lisboa, Portugal\\
$^{5}$Laboratoire Kastler Brossel, École Normale Supérieure,
CNRS, Université P. et M. Curie -- Paris 6, Case 74; 4, place
Jussieu, 75252 Paris CEDEX 05, France}

\date{\today}

%
%
%
%

\begin{abstract}
Two--photon decay of hydrogen--like ions is studied within the
framework of second--order perturbation theory, based on
relativistic Dirac's equation. Special attention is paid to the
effects arising from the summation over the negative--energy
(intermediate virtual) states that occurs in such a framework. In
order to investigate the role of these states, detailed calculations
have been carried out for the $2s_{1/2} \to 1s_{1/2}$ and $2p_{1/2}
\to 1s_{1/2}$ transitions in neutral hydrogen H as well as for
hydrogen--like xenon Xe$^{53+}$ and uranium U$^{91+}$ ions. We found
that for a correct evaluation of the total and energy--differential
decay rates, summation over the negative--energy part of Dirac's
spectrum should be properly taken into account both for high--$Z$
and low--$Z$ atomic systems.
\end{abstract}


\maketitle

%
%
%
%

%
%
\section{Introduction}

Experimental and theoretical studies on the two--photon transitions
in atomic systems have a long tradition. Following seminal works by
Göppert--Mayer \cite{GoM31} and by Breit and Teller \cite{BrT40} a
large number of investigations have been performed in the past which
focused on the decay of metastable states of light neutral atoms and
low--$Z$ ions. These investigations have dealt not only with the
total and energy--differential decay rates \cite{SpG51,ShB59,Flo84}
but also with the angular distributions
\cite{MaS72,LiN65,Kla69,CAu76} and even polarization correlations
between the two emitted photons \cite{CoK75,AsD82,KlD97}. Detailed
analysis of these two--photon properties have revealed unique
information about electron densities in astrophysical plasmas and
thermal X--ray sources, highly precise values of physical constants
\cite{ScJ99}, structural properties of few--electron systems
including subtle quantum electrodynamical (QED) effects \cite{GoM83}
as well as about the basic concepts of quantum physics such as,
e.g., non--locality and non--separability \cite{RaS08}.

Beside the decay of metastable states of low--$Z$ systems, much of
today's interest is focused also on the two--photon transitions in
high--$Z$ ions and atoms which provide a sensitive tool for
improving our understanding of the electron--photon interactions in
the presence of extremely strong electromagnetic fields
\cite{FrI05}. In such strong fields produced by heavy nuclei,
relativistic and retardation effects become of paramount importance
and may strongly affect the properties of two--photon emission. To
explore these effects, therefore, theoretical investigations based
on Dirac's equation have been carried for the total and
energy--differential decay rates
\cite{GoD81,DrG81,SaP98,JeS08,AmS09} as well as for the angular and
polarization correlations \cite{ToL90,SuK05,SuR09}. In general,
relativistic predictions for the two--photon total and differential
properties have been found in a good agreement with experimental
data obtained for the decay of inner--shell vacancies of heavy
neutral atoms \cite{MoD04,IlU06} and excited states of high--$Z$
few--electron ions \cite{KuT09}.

Although intensive experimental and theoretical efforts have been
undertaken recently to understand relativistic effects on the
two--photon transitions in heavy ions and atoms, a number of
questions still remain open. One of the questions, which currently
attracts much of interest, concerns the role of \textit{negative
energy} solutions of Dirac's equation in relativistic two--photon
calculations. Usually, these calculations are performed within the
framework of the second--order perturbation theory and, hence,
require summation over the (virtual) intermediate ion states. Such a
summation, running over the \textit{complete} spectrum, should
obviously include not only positive-- (discrete and continuum) but
also negative--eigenenergy Dirac's states. One might expect,
however, that since the energy release in two--photon bound--bound
transitions is less than the energy required for the
electron--positron pair production, the contribution from the
negative part of Dirac's spectrum should be negligible even for the
decay of heaviest elements. From practical viewpoint, this
assumption justifies the restriction of the intermediate--state
summation to the positive--energy solutions only. Exclusion of the
negative continuum would lead, in turn, to a significant
simplification of the the second--order relativistic calculations
especially for many--electron systems for which the problem of (many
particle) negative continuum still remains unsolved.

Despite the (relatively) small energy of two--photon transitions,
the influence of Dirac's negative continuum in second--order
calculations should be further questioned because of possibility for
production and subsequent annihilation of the \textit{virtual}
anti--particles. It has been argued, for example, that transitions
involving positron states have to be taken into account for the
proper description of Thomson scattering \cite{Sak67}, interaction
of ions with intense electromagnetic pulses \cite{BoF08,SeL09} in
the ``undercritical'' regime as well as magnetic transitions in
two--electron ions \cite{LiS90,DeS98,Ind96}. Moreover, the first
step towards the analysis of negative--energy contributions to the
two--photon properties has been done by Labzowsky and co--workers
\cite{LaS05} who focused on E1M1 and E1E2 $2p_{1/2} \to 1s_{1/2}$
total decay probabilities. The relativistic calculations have
indicated the importance of negative--energy contributions not only
for high--$Z$ but also for low--$Z$ hydrogen--like ions.

In this work, we apply the second--order perturbation theory based
on relativistic Dirac's equation in order to re--analyze atomic
two--photon decay. We pay special attention to the influence of
negative continuum solutions on the evaluation of the transition
amplitudes and, hence, on the total and energy--differential decay
rates. For the sake of clarity, we restrict our analysis to the
decay of hydrogen--like ions for which both the positive-- and
negative--energy parts of Dirac's spectrum can be still studied in a
systematic way by making use of a finite basis set method
\cite{SaP98}. Implementation of this method for computing
relativistic second--order transition amplitudes is briefly
discussed in Sections \ref{sub_sec_theory} and
\ref{sub_sec_Bspline}. Later, in Section
\ref{sub_sect_semi_relativistic}, we consider an alternative,
semi--classical, approach which allows \textit{analytical}
evaluation of the negative--energy contributions to the two--photon
matrix elements and transition rates. These two---semi--classical
and fully relativistic---approaches are used in Section
\ref{Sec_results} to calculate the energy--differential and total
decay rates for several multipole terms in the $2s_{1/2} \to
1s_{1/2}$ and $2p_{1/2} \to 1s_{1/2}$ two--photon decay of neutral
hydrogen as well as hydrogen--like xenon Xe$^{53+}$ and uranium
U$^{91+}$ ions. Based on the results of our calculations, we argue
that both the total transition probabilities and the photon energy
distributions can be strongly affected by the negative--state
contributions; this effect is most clearly observed for the
non--dipole transitions not only in high--$Z$ but also in
(non--relativistic) low--$Z$ domain. Brief summary of these findings
and outlooks are given finally in Section \ref{Sec_summary}.

%
%

\section{Theory}
\label{Sec_theory}

\subsection{Differential and total decay rates}
\label{sub_sec_theory}

Not much has to be said about the basic formalism for studying the
two--photon transitions in hydrogen--like ions. In the past, this
formalism has been widely applied in order to investigate not only
the total decay probabilities \cite{DrG81, GoD81, SaP98, LaS05} but
also the energy as well as angular distributions \cite{SuK05} and
even the correlation in the polarization state of the photons
\cite{RaS08,SuR09}. Below, therefore, we restrict ourselves to a
rather brief account of the basic expressions, just enough for
discussing the role of negative--energy solutions of Dirac's
equation in computing of the two--photon (total and differential)
rates.

The properties of the two--photon atomic transitions are evaluated,
usually, within the framework of the second--order perturbation
theory. When based on Dirac's equation, this theory gives the
following expression for the differential in energy decay rate:
\begin{eqnarray}
   \label{rate_general}
   \frac{{\rm d}w}{{\rm d}\omega_1} &=& \frac{\omega_1 \omega_2}{(2 \pi)^3 c^3} \,
   \Bigg| \sum\limits_{\nu} \Bigg( \frac{\mem{f}{{\bm A}_2^*}{\nu}
   \mem{\nu}{{\bm A}_1^*}{i}}{E_{\nu} - E_i + \omega_1} \nonumber \\
   &+& \frac{\mem{f}{{\bm A}_1^*}{\nu}
   \mem{\nu}{{\bm A}_2^*}{i}}{E_{\nu} - E_i + \omega_2} \Bigg) \Bigg|^2 \,
   {\rm d}\Omega_1 {\rm d}\Omega_2 \, ,
\end{eqnarray}
where the transition operators ${\bm A}^*_{j}$ with $j = 1, 2$
describe the (relativistic) electron--photon interaction. For the
emission of photons with wave vectors ${\bm k}_{j}$ and polarization
vectors $\hat{{\bm e}}_{j}$ these operators read as:
\begin{equation}
   \label{A_operator_general}
   {\bm A}^*_{j} = {\bm \alpha} \cdot \left( \hat{{\bm e}}_{j} + G {\bm k}_{j} \right)
   {\rm e}^{-i {\bm k}_{j} {\bm r}} - G {\rm e}^{-i {\bm k}_{j} {\bm r}} \, ,
\end{equation}
where ${\bm \alpha}$ is a vector of Dirac matrices and $G$ is an
arbitrary gauge parameter. In the calculations below, following
Grant \cite{Gra74}, we employ two different gauges that are known to
lead to well known non--relativistic operators. First, we use the
so--called Coulomb gauge, when $G$ =0, which corresponds to the
velocity form of electron--photon interaction operator in the
non--relativistic limit. As the second choice we adopt $G =
\sqrt{(L+1)/L}$ in order to obtain Babushkin gauge which reduces,
for the particular case of $L$=1, to the dipole length form of the
transition operator.

In Eq.~(\ref{rate_general}), $\ketm{i} \equiv \ketm{n_i \kappa_i
\mu_i}$ and $\ketm{f} \equiv \ketm{n_f \kappa_f \mu_f}$ denote
solutions of the Dirac's equation for the initial and final ionic
states while $E_i \equiv E_{n_i \kappa_i}$ and $E_f \equiv E_{n_f
\kappa_f}$ are the corresponding one--particle energies. Because of
energy conservation, $E_i$ and $E_f$ are related to the energies
$\omega_{1,2} = c k_{1,2}$ of the emitted photons by:
\begin{equation}
   \label{energy_conservation}
   E_i - E_f = \omega_1 + \omega_2 \, .
\end{equation}
>From this relation, it is convenient to define the so-called energy
sharing parameter $y = \omega_1/(\omega_1 + \omega_2)$, i.e., the
fraction of the energy which is carried away by the ``first'' photon.

As usual in atomic physics, the second--order transition amplitudes
in Eq. (\ref{rate_general}) and, hence, the two--photon transitions
rates can be further simplified by applying the techniques of
Racah's algebra if all the operators are presented in terms of
spherical tensors and if the (standard) radial--angular
representation of Dirac's wavefunctions are employed. For the
interaction of electron with electromagnetic field, the spherical
tensor components are obtained from the \textit{multipole} expansion
of the operator ${\bm A}^*_{j}$ (see Refs. \cite{GoD81,SaP98,VaM88}
for further details). By using such an expansion, we are able to
re--write Eq. (\ref{rate_general}) as a sum of partial multipole
rates
\begin{eqnarray}
   \label{rate_expansion}
   \frac{{\rm d}w}{{\rm d}\omega_1} &=& \sum\limits_{\Theta_1 L_1 \Theta_2 L_2} \,
   \frac{{\rm d}W_{\Theta_1 L_1 \Theta_2 L_2}}{{\rm d}\omega_1} \, ,
\end{eqnarray}
which describe the emission of two photons of electric ($\Theta_j$ =
E) or/and magnetic ($\Theta_j$ = M) type carrying away the angular
momenta $L_1$ and $L_2$. For the decay of unpolarized ionic state
$\ketm{n_i \kappa_i}$, in which the emission angles as well as
polarization of both photons remain unobserved, these partial
multipole rates are given by \cite{GoD81}:
\begin{eqnarray}
   \label{rate_partial}
   \frac{{\rm d}W_{\Theta_1 L_1 \Theta_2 L_2}}{{\rm d}\omega_1} &=& \frac{\omega_1 \omega_2}{(2 \pi)^3 c^3} \,
   \sum\limits_{\lambda_{\Theta_1} \lambda_{\Theta_2}} \, \sum\limits_{\kappa_\nu} \, \Bigg[
   \left| S^{j_\nu}(1, 2) \right|^2 \nonumber \\
   && \hspace*{-2.7cm} + \left| S^{j_\nu}(2, 1) \right|^2 \, + \, 2 \sum\limits_{\kappa_\nu'} d(j_\nu, j_\nu')
   \, S^{j_\nu}(2, 1) S^{j_\nu'}(1, 2) \Bigg] \, ,
\end{eqnarray}
where the angular coefficient $d(j_\nu, j_\nu')$ is defined by the phase factor and 6j Wigner symbol:
\begin{eqnarray}
   \label{d_definition}
   d(j_\nu, j_\nu') &=& \sqrt{(2j_\nu + 1)(2j_\nu' + 1)} \nonumber \\
   &×& (-1)^{2j_\nu + L_1 + L+2} \, \sixjm{j_f}{j_\nu'}{L_1}{j_i}{j_\nu}{L_2} \, ,
\end{eqnarray}
and the radial integral part is expressed in terms of the reduced
matrix elements of the multipole (electric and magnetic) field
operators:
\begin{eqnarray}
   \label{S_function_def}
   S^{j_\nu}(1, 2) & & \nonumber \\
   && \hspace*{-2.0cm} = \sum\limits_{n_\nu}
   \frac{\rmem{n_f \kappa_f}{\hat{a}^{\lambda_{\Theta_1} *}_{L_1}}{n_\nu \kappa_\nu}
   \rmem{n_\nu \kappa_\nu}{\hat{a}^{\lambda_{\Theta_2} *}_{L_2}}{n_i \kappa_i}}{E_{\nu} - E_i + \omega_2} \, .
\end{eqnarray}
The summation over $\lambda_{\Theta_j}$ in Eq.~(\ref{rate_partial})
is restricted to $\lambda_{\Theta_j} = ± 1$ for the electric
($\Theta_j$ = E) and $\lambda_{\Theta_j} = 0$ for the  magnetic
($\Theta_j$ = M) photon transitions.

Until now, we have discussed the general expressions for the
two--photon transition rates which are differential in energy
$\omega_1$ of one of the photons. By performing an integration over
this energy one may easily obtain the total rate that is directly
related to the lifetime of a particular excited state against the
two--photon decay. As it follows from Eq.~(\ref{rate_expansion}),
such a total rate can be represented as a sum of its multipole
components:
\begin{eqnarray}
   \label{total_rate}
   w_{\rm tot} &=& \sum\limits_{\Theta_1 L_1 \Theta_2 L_2}
   W_{\Theta_1 L_1 \Theta_2 L_2} \nonumber \\
   &\equiv&  \sum\limits_{\Theta_1 L_1 \Theta_2 L_2}
   \int\limits_{0}^{\omega_t}{\frac{{\rm d}W_{\Theta_1 L_1 \Theta_2 L_2}}{{\rm d}\omega_1} \,
   {\rm d}\omega_1} \, ,
\end{eqnarray}
where $\omega_t = E_i - E_f$ is the transition energy.

As seen from Eqs.~(\ref{rate_expansion})--(\ref{total_rate}), any
analysis of the differential as well as total two--photon decay
rates can be traced back to the (reduced) matrix elements that
describe the interaction of an electron with the (multipole)
radiation field. Since the relativistic form of these matrix
elements is applied very frequently in studying the various atomic
processes, we shall not discuss here their evaluation and just refer
the reader for all details to references \cite{GoD81,SaP98,Gra74}.
Instead, in the next section we will focus on the summation over the
intermediate states $\ketm{n_\nu \kappa_\nu}$ which appears in the
second--order transition amplitudes (see
Eq.~(\ref{S_function_def})).

\begin{figure}
\vspace*{0.0cm}
\includegraphics[height=11.0cm,angle=00,clip=]{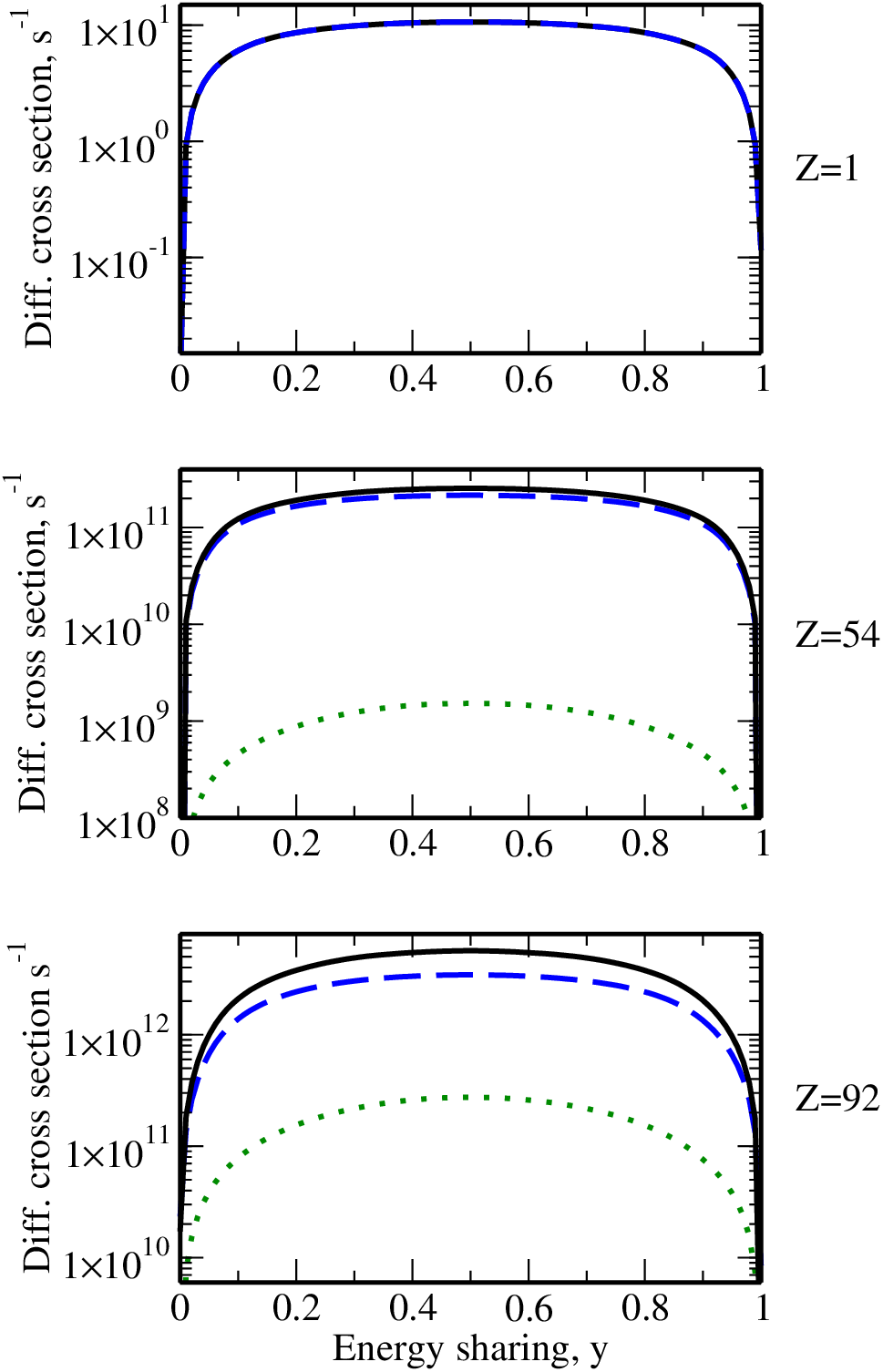}
\vspace*{0.0cm} \caption{(Color online) Energy--differential
transition rates for the 2E1 $2s_{1/2} \to 1s_{1/2}$ two--photon
decay of hydrogen and hydrogen--like ions. Relativistic calculations
have been carried out by performing intermediate--state summation
over complete Dirac's spectrum (solid line) as well as by
restricting this summation to the positive-- (dashed line) and
negative--energy (dotted line) states only.} \label{Fig1}
\end{figure}

\subsection{Summation over the intermediate states}
\label{sub_sec_Bspline}

The summation over the intermediate states in
Eq.~(\ref{S_function_def}) runs over the complete one--particle
spectrum $\ketm{n_\nu \kappa_\nu}$, including a summation over the
discrete part of the spectrum as well as an integration over the
positive and negative--energy continuum. In practice, of course,
performing such an infinite--state summation is a rather demanding
task. A number of methods have been developed over the last decades
in order to evaluate the second--order transition amplitudes
consistently. Apart from the Green's function approach
\cite{SuK05,SwD91} which ---in case of a purely Coulomb potential---
allows for the analytical computation of Eq.~(\ref{S_function_def}),
the \textit{discrete--basis--set} summation is widely used nowadays
in two--photon studies \cite{SaP98}. A great advantage of the latter
method is that it allows to separate the contributions from the
positive-- and negative--energy solutions in the intermediate--state
summation. Since the effects that arise from the negative--energy
spectrum are in the focus of the present study, we apply for the
calculations below the finite (discrete) basis solutions constructed
from the B--spline sets.

Although the B--spline basis set approach has been discussed in
detail elsewhere \cite{SaP98,JoB88,LaG97}, here we briefly recall
its main features. In this way, we shall consider the ion (or atom)
under consideration to be enclosed in a finite cavity with a radius
$R$ large enough to get a good approximation of the wavefunctions
with some suitable set of boundary conditions, which allows for
discretization of the continua. Wavefunctions that describe the
quantum states $\ketm{\nu} \equiv \ketm{n_\nu \kappa_\nu}$ of such a
``particle in box'' system can be expanded in terms of basis set
functions $\phi_{\nu}^i(r)$ with $i = 1, .. , 2N$ which, in turn,
are found as solutions of the Dirac--Fock equation,
\begin{eqnarray}
   \label{dirac-fock-eq}
   \left[ \begin{array}{cc}
                    \frac{V(r)}{c} & \frac{{\rm d}}{{\rm d}r} - \frac{\kappa_{\nu}}{R}  \\[0.2cm]
                    - \left( \frac{{\rm d}}{{\rm d}r} + \frac{\kappa_{\nu}}{R} \right) & -2c + \frac{V(r)}{c}
          \end{array} \right] \phi_{\nu}^i(r) = \frac{\epsilon^i_\nu}{c} \phi_{\nu}^i(r) \, ,
\end{eqnarray}
where $\epsilon_\nu^i = E_\nu^i - mc^2$ and $V(r)$ is a Coulomb
potential of a uniformly charged finite--size nucleus. Due to
computational reasons, each of $\phi_{\nu}^i(r)$ function is
expressed as a linear combination of B--splines as it was originally
proposed in Ref.~\cite{JoB88} by Johnson and co--workers.

For each quantum state $\ketm{\nu}$ the set of basis functions
$\phi_{\nu}^i(r)$ spans both positive and negative energy solutions.
Solutions labeled by $i = 1, .., N$ describe the negative continuum
with $\epsilon_\nu^i < -2mc^2$ while solutions labeled by $i = N+1,
.., 2N$ correspond to the first few states of the bound--state
spectrum as well as to positive continuum with $\epsilon_\nu^i > 0$.
Thus, by selecting the proper sub--set of basis functions
$\phi_{\nu}^i(r)$ we may explore the role of negative continuum in
computing of the properties of two--photon emission from
hydrogen--like ions.

\subsection{Semi--relativistic approximation}
\label{sub_sect_semi_relativistic}

Based on the relativistic theory, the expressions obtained in the
previous section allow to study the influence of the Dirac's
negative continuum on the properties of two--photon emission from
hydrogen--like ions with nuclear charge in the whole range $1 \leq Z
\leq 92$. For the low--$Z$ ions, moreover, it is also useful to
estimate the negative--energy contributions within the
semi--relativistic approach as proposed in the work by Labzowsky and
co--workers \cite{LaS05}. To perform such a semi--relativistic
analysis let us start from Eq.~(\ref{rate_general}) in which we
retain the sum only over the negative--energy continuum states.
Since the total energy of these states is $E_{\nu} = - (T_{\nu} +
mc^2)$, the corresponding energy denominator of the second--order
transition amplitude can be written as $E_{\nu} - E_i + \omega_j
\approx - 2mc^2$ which leads to the following expression for the
differential decay rate:
\begin{eqnarray}
   \label{rate_general_nr_negative}
   \frac{{\rm d}w^{(-)}}{{\rm d}\omega_1} &=& \frac{\omega_1 \omega_2}{(2 \pi)^3 c^3} \,
   \frac{1}{4 (mc^2)^2} \,
   \Bigg| \sum\limits_{\nu \in (-)} \Big(
   \mem{f}{{\bm A}_2^*}{\nu}
   \mem{\nu}{{\bm A}_1^*}{i}
   \nonumber \\
   &+&
   \mem{f}{{\bm A}_1^*}{\nu}
   \mem{\nu}{{\bm A}_2^*}{i}
   \Big) \Bigg|^2 \,
   {\rm d}\Omega_1 {\rm d}\Omega_2 \, .
\end{eqnarray}
For the further simplification of this expression we shall make use
of the multipole expansion of the electron--photon interaction
operators (\ref{A_operator_general}). For the sake of simplicity, we
restrict this semi--relativistic analysis to the case of Coulomb
gauge ($G$ = 0) in which operator ${\bm A}^*_{j}$ can be written as:
\begin{equation}
   \label{A_operator_decomposition}
   {\bm A}^*_{j} = {\bm \alpha} \cdot \hat{{\bm e}}_{j} \,
   (1 - i {\bm k} \cdot {\bm r} +
   1/2 \left( - i {\bm k} \cdot {\bm r} \right)^2 + ... ) \, ,
\end{equation}
if one expand the photon exponential ${\rm exp}(i {\bm k} \cdot {\bm
r})$ into the Taylor series.

\begin{figure}
\vspace*{0.0cm}
\includegraphics[height=11.0cm,angle=00,clip=]{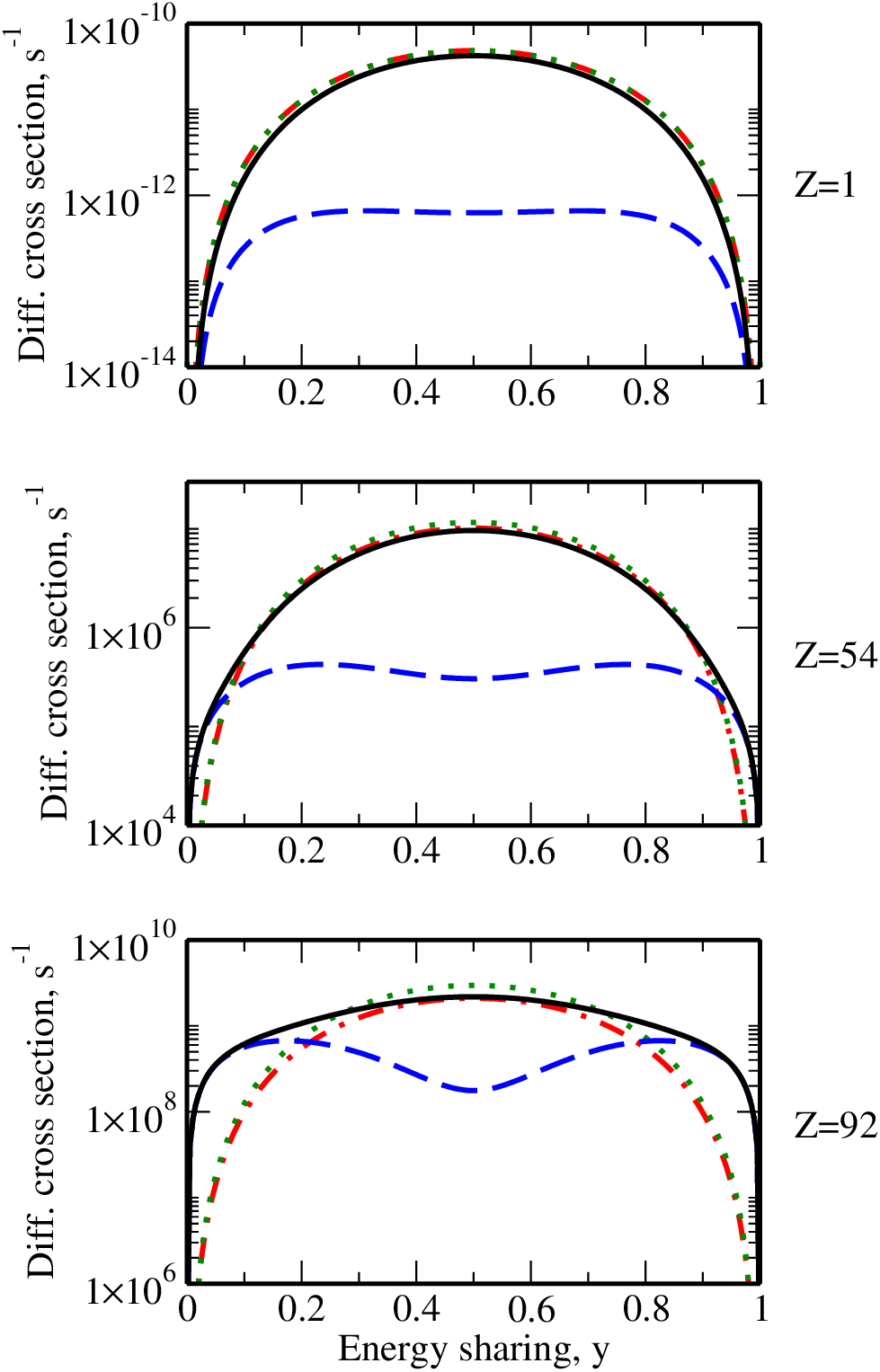}
\vspace*{0.0cm} \caption{(Color online) Energy--differential decay
rates for the (sum of the) 2M1, 2E2 and E2M1 $2s_{1/2} \to 1s_{1/2}$
multipole two--photon transitions in hydrogen and hydrogen--like
ions. Relativistic calculations have been carried out by performing
intermediate--state summation over complete Dirac's spectrum (solid
line) as well as by restricting this summation to the positive--
(dashed line) and negative--energy (dotted line) states only.
Results of relativistic calculations are compared also with the
semi--relativistic prediction (dot--dashed line) as given by
Eq.~(\ref{rate_general_nr_negative_M1M1_E2E2_3}).} \label{Fig2}
\end{figure}

In contrast to the ``standard'' spherical tensor expansion
\cite{GoD81,VaM88}, the series (\ref{A_operator_decomposition})
usually does not allow one to make a clear distinction between the
different multipole components of the electromagnetic field. For
instance, while the first term in
Eq.~(\ref{A_operator_decomposition}) describes---within the
non--relativistic limit---electric dipole (E1) transition, the term
$\left(- i {\bm k} \cdot {\bm r} \right)$ gives rise both, to
magnetic dipole (M1) and electric quadrupole (E2) channels. Such an
approximation, however, is well justified for our
(semi--relativistic) analysis which just aims to estimate the role
of negative continuum states in the different (\textit{groups of})
multipole two--photon transitions in light hydrogen--like ions. In
particular, by adopting ${\bm A}^*_{j} = - {\bm \alpha} \cdot
\hat{{\bm e}}_{j} \, (i {\bm k} \cdot {\bm r})$ for both operators
in Eq.~(\ref{rate_general_nr_negative}) we may find the contribution
from the negative spectrum to the 2M1, 2E2 and E2M1 $2s_{1/2} \to
1s_{1/2}$ transition probabilities:
\begin{eqnarray}
   \label{rate_general_nr_negative_M1M1_E2E2}
   \frac{{\rm d}w^{(-)}_{\rm M1, E2}}{{\rm d}\omega_1} &=& \frac{\omega_1
   \omega_2}{(2 \pi)^3 c^3} \, \frac{1}{4 (mc^2)^2} \nonumber \\
   && \hspace*{-3cm} ×
   \Bigg| \sum\limits_{\nu \in (-)} \Big(
   \mem{f}{{\bm \alpha} \cdot
   \hat{{\bm e}}_{2} \, ({\bm k}_2 \cdot {\bm r})}{\nu}
   \mem{\nu}{{\bm \alpha} \cdot
   \hat{{\bm e}}_{1} \, ({\bm k}_1 \cdot {\bm r})}{i}
   \nonumber \\
   && \hspace*{-3cm} +
   \mem{f}{{\bm \alpha} \cdot
   \hat{{\bm e}}_{1} \, ({\bm k}_1 \cdot {\bm r})}{\nu}
   \mem{\nu}{{\bm \alpha} \cdot
   \hat{{\bm e}}_{2} \, ({\bm k}_2 \cdot {\bm r})}{i}
   \Big) \Bigg|^2 \, \nonumber \\
   && \hspace*{-3cm} ×
   {\rm d}\Omega_1 {\rm d}\Omega_2 \, .
\end{eqnarray}
Here, summation over the intermediate states $\ketm{\nu}$ is
restricted by the negative--energy solutions of the Dirac equation
for the electron in the field of nucleus. In the non--relativistic
limits these states form a \textit{complete} set of solutions of the
Schrödinger equation for the particle in a repulsive Coulomb filed
\cite{LaL77}. By employing a closure relation for such a set we
re--write Eq.~(\ref{rate_general_nr_negative_M1M1_E2E2}) in the
form:
\begin{eqnarray}
   \label{rate_general_nr_negative_M1M1_E2E2_2}
   \frac{{\rm d}w^{(-)}_{\rm M1, E2}}{{\rm d}\omega_1} &=& \frac{\omega_1
   \omega_2}{(2 \pi)^3 c^3} \, \frac{1}{(mc^2)^2} \nonumber \\
   && \hspace*{-2cm} ×
   \left|( \hat{{\bm e}}_{1} \hat{{\bm e}}_{2} )
   \mem{f}{({\bm k}_1 \cdot {\bm r}) ({\bm k}_2 \cdot {\bm r})}{i}
   \right|^2 \, {\rm d}\Omega_1 {\rm d}\Omega_2 \, ,
\end{eqnarray}
where $\ketm{i}$ and $\ketm{f}$ denote now the solutions of the
Schrödinger equation for the initial and final ionic states,
respectively. For the particular case of $2s_{1/2} \to 1s_{1/2}$
two--photon transition, i.e., when $\ketm{i} = \ketm{2s}$ and
$\ketm{f} = \ketm{1s}$, this expression finally reads:
\begin{eqnarray}
   \label{rate_general_nr_negative_M1M1_E2E2_3}
   \frac{{\rm d}w^{(-)}_{\rm M1, E2}}{{\rm d}\omega_1} &=&
   \frac{2^{22}}{3^{13}} \, \frac{\alpha^{10}}{5 \pi Z^4} \,
   \omega_1^3 \omega_2^3 \, ,
\end{eqnarray}
if one performs an integration over the photon emission angles as
well as a summation over the polarization states (see
Ref.~\cite{LaS05} for further details).

Eq.~(\ref{rate_general_nr_negative_M1M1_E2E2_3}) provides the
differential rate for the 2M1, 2E2 and E2M1 two--photon transitions
as obtained within the non--relativistic framework and by
restricting the summation over the intermediate spectrum
$\ketm{\nu}$ to the negative energy states only. Being valid for
low--$Z$ ions, this expression may also help us to analyze the
negative--energy contribution to the \textit{total} decay rate,
\begin{eqnarray}
   \label{total_nr_negative_M1M1_E2E2}
   w^{(-)}_{\rm M1, E2} &=&
   \int{\frac{{\rm d}w^{(-)}_{M1, E2}}{{\rm d}\omega_1} \, {\rm d}\omega_1}
   = (\alpha Z)^{10} \, \frac{1}{14 \, \pi \, 5^2 \, 3^6} \nonumber \\
   &=& 1.247
   \, 10^{-6} \,(\alpha Z)^{10} \, ,
\end{eqnarray}
where the integration over the photon energy $\omega_1$ is
performed.

\begin{figure}
\vspace*{0.0cm}
\includegraphics[height=11.0cm,angle=00,clip=]{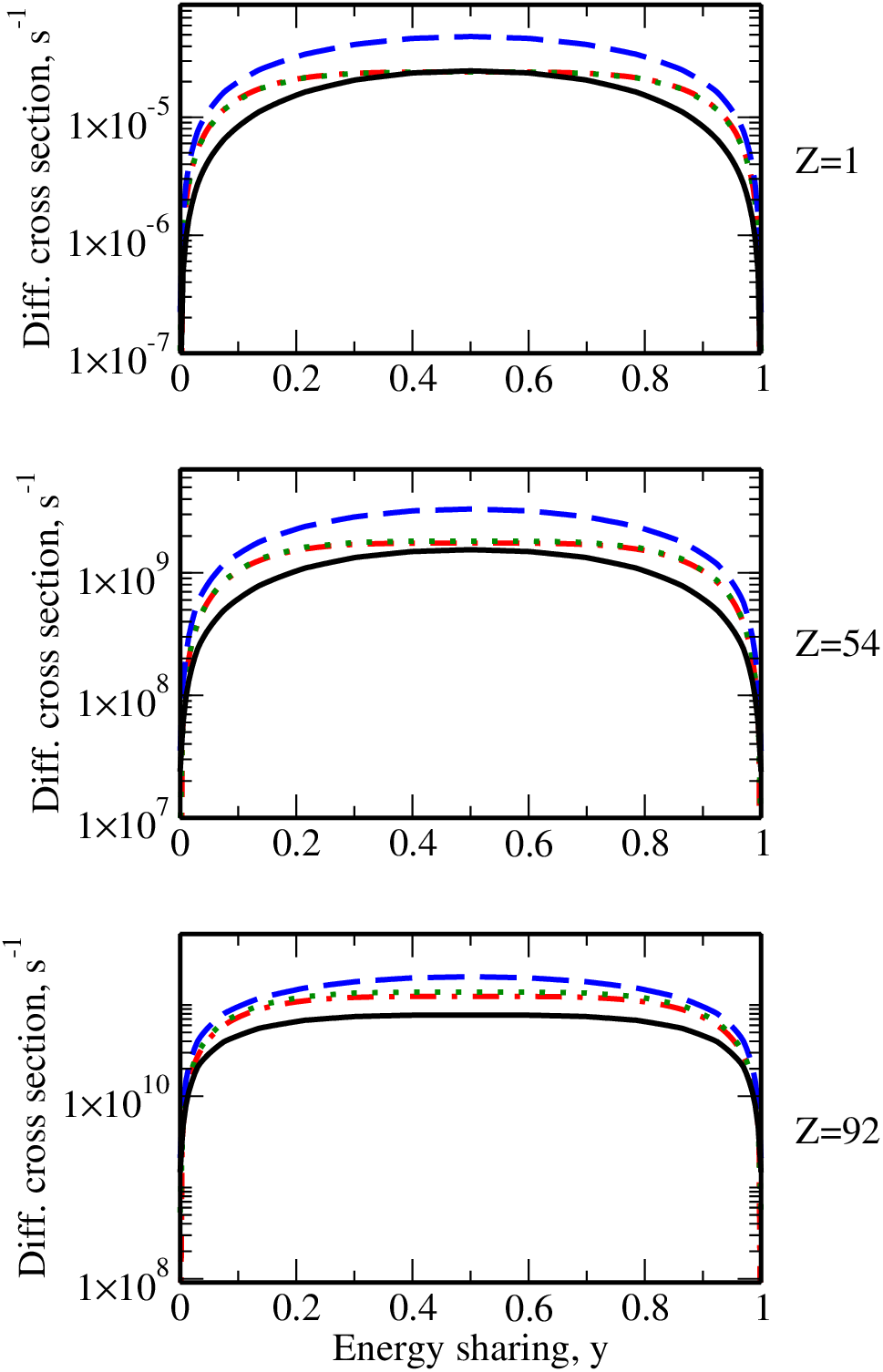}
\vspace*{0.0cm} \caption{(Color online) Energy--differential decay
rates for the (sum of the) E1M1 and E1E2 $2p_{1/2} \to 1s_{1/2}$
multipole two--photon transitions in hydrogen and hydrogen--like
ions. Relativistic calculations have been carried out by performing
intermediate--state summation over complete Dirac's spectrum (solid
line) as well as by restricting this summation to the positive--
(dashed line) and negative--energy (dotted line) states only.
Results of relativistic calculations are compared also with the
semi--relativistic prediction (dot--dashed line) as given by
Eq.~(\ref{rate_general_nr_negative_E1M1_E1E2}).} \label{Fig3}
\end{figure}

Apart from the 2M1, 2E2 and E2M1 $2s_{1/2} \to 1s_{1/2}$ two--photon
transitions, Eqs.~(\ref{rate_general_nr_negative}) and
(\ref{A_operator_decomposition}) may also be employed to study other
decay channels. For example, the negative energy contributions to
the differential as well as total rates for the E1M1 and E1E2
$2p_{1/2} \to 1s_{1/2}$ decay read as:
\begin{eqnarray}
   \label{rate_general_nr_negative_E1M1_E1E2}
   \frac{{\rm d}w^{(-)}_{\rm E1, M1, E2}}{{\rm d}\omega_1} &=&
   \frac{2^{17}}{3^{12}} \, \frac{\alpha^8}{\pi Z^2} \,
   \omega_1 \omega_2 (\omega_1^2 + \omega_2^2) \, ,
\end{eqnarray}
and
\begin{eqnarray}
   \label{total_nr_negative_E1M1_E1E2}
   w^{(-)}_{\rm E1, M1, E2}
   &=& (\alpha Z)^{8} \, \frac{2}{5 \pi \, 3^{7}} \nonumber \\
   &=& 5.822
   \, 10^{-5} \,(\alpha Z)^{8} \, ,
\end{eqnarray}
respectively \cite{LaS05}. Together with
Eqs.~(\ref{rate_general_nr_negative_M1M1_E2E2_3}) and
(\ref{total_nr_negative_M1M1_E2E2}), we shall later use these
non--relativistic predictions in order to check the validity of our
numerical calculations in low--$Z$ domain.

%
%
%
%
%
\begin{table*}[t]
\begin{center}
\begin{ruledtabular}
\begin{tabular}{|l l|l|l|l|l|l|l|}
 & & \multicolumn{2}{c|}{$Z$=1} & \multicolumn{2}{c|}{$Z$=54} & \multicolumn{2}{c|}{$Z$=92} \\
\cline{3-8}
 & & length & velocity & length & velocity & length & velocity\\
\hline
       & W$_{+}$     & 8.2291 (+00)  & 8.2291 (+00)  &
                       1.6311 (+11) & 1.6023 (+11) &
                       2.9041 (+12) & 2.3939 (+12) \\
2E1   & W$_{-}$     & 2.4949 (-08)  & 6.2372 (-09)  &
                       3.8442 (+09)  & 9.6290 (+08)  &
                       6.8066 (+11) & 1.7044 (+11) \\
       & W$_{\rm t}$ & 8.2291 (+00)  & 8.2291 (+00)  &
                       1.8592 (+11) & 1.8592 (+11) &
                       3.8256 (+12) & 3.8256 (+12) \\
\hline
       & W$_{+}$     & 2.5372 (-10) & 2.5372 (-10) &
                       4.7949 (+07) & 4.7940 (+07) &
                       8.2955 (+09) & 8.2714 (+09)  \\
E1M2   & W$_{-}$     & 9.1743 (-21) & 4.5871 (-21) &
                       1.9521 (+04) & 9.7905 (+03) &
                       4.8084 (+07) & 2.4070 (+07) \\
       & W$_{\rm t}$ & 2.5372 (-10) & 2.5372 (-10) &
                       4.9278 (+07) & 4.9278 (+07) &
                       9.1387 (+09) & 9.1387 (+09) \\
\hline
       & W$_{+}$     & 3.7296 (-11) & 4.8617 (-13) &
                       9.1765 (+06)  & 1.9624 (+05)  &
                       2.4730 (+09) & 9.7383 (+07) \\
2E2   & W$_{-}$     & 4.5092 (-11) & 8.2822 (-12) &
                       1.1000 (+07) & 2.0202 (+06) &
                       2.9087 (+09) & 5.3305 (+08) \\
       & W$_{\rm t}$ & 4.9072 (-12) & 4.9072 (-12) &
                       9.8177 (+05) & 9.8177 (+05)  &
                       1.7859 (+08) & 1.7859 (+08) \\
\hline
       & W$_{+}$     & \multicolumn{2}{c|}{5.9021 (-20)} &
                       \multicolumn{2}{c|}{1.2691 (+05)} &
                       \multicolumn{2}{c|}{3.3321 (+08)} \\
2M1   & W$_{-}$     & \multicolumn{2}{c|}{1.3804 (-11)} &
                       \multicolumn{2}{c|}{3.2695 (+06)} &
                       \multicolumn{2}{c|}{7.9720 (+08)}  \\
       & W$_{\rm t}$ & \multicolumn{2}{c|}{1.3804 (-11)} &
                       \multicolumn{2}{c|}{3.4027 (+06)} &
                       \multicolumn{2}{c|}{1.1093 (+09)}  \\
\end{tabular}
\end{ruledtabular}
\caption{Total rates (in s$^{-1}$) for the several multipole
combinations of $2s_{1/2} \to 1s_{1/2}$ two--photon decay.
Relativistic calculations have been performed within the velocity
and length gauges and by carrying out intermediate--state summation
over the complete Dirac's spectrum (W$_{\rm t}$) as well as over the
positive-- (W$_{+}$) and negative--energy (W$_{-}$) solutions only.}
\end{center}
\vspace{-0.6cm}
\end{table*}
%
%
%
%
%

%
%

\section{Results and discussion}
\label{Sec_results}

Having discussed the theoretical background for the two--photon
studies, we are prepared now to analyze the influence of the Dirac's
negative continuum on the total as well as energy--differential
decay rates. We shall start such an analysis from the $2s_{1/2} \to
1s_{1/2}$ transition, which is well established both in theory
\cite{GoD81,SaP98,SuK05} and in experiment. For all hydrogen--like
ions this transition is dominated by the 2E1 decay channel while all
the higher multipoles contribute by less than 0.5\% to the decay
probability. The energy--differential decay rate given by
Eq.~(\ref{rate_partial}) for the emission of two electric dipole
photons is displayed in Fig.~\ref{Fig1} for the decay of neutral
hydrogen (H) as well as hydrogen--like xenon Xe$^{53+}$ and uranium
U$^{91+}$ ions. For these ions, relativistic second--order
calculations have been done within the Coulomb gauge and by
performing intermediate--state summation over the complete Dirac's
spectrum (solid line) as well as over the positive-- (dashed line)
and negative--energy (dotted line) solutions only. As seen from the
figure, the negative--energy contribution to the
energy--differential decay rate is negligible for low--$Z$ ions but
becomes rather pronounced as the nuclear charge $Z$ is increased.
For the 2E1 decay of hydrogen--like uranium, for example, exclusion
of the negative solutions from the intermediate--state summation in
Eq.~(\ref{S_function_def}) leads to about 20 \% reduction of the
decay rate when compared with the ``exact'' result.

While for the leading, 2E1 $2s_{1/2} \to 1s_{1/2}$ transition the
negative continuum effects arise only for rather heavy ions, they
might strongly affect properties of the higher multipole decay
channels in low--$Z$ domain. In Fig.~\ref{Fig2}, for example, we
display the energy distributions of photons emitted in 2M1 and 2E2
transitions. As seen from the upper panel of the figure
corresponding to the decay of neutral hydrogen, negative energy part
of the Dirac's spectrum gives the dominant contribution to the (sum
of the) differential rates for these decay channels. With the
increasing nuclear charge $Z$, the role of positive energy solutions
also becomes more pronounced. However, these solutions allow one to
describe reasonably well the differential rates (\ref{rate_partial})
only if one of the photons is much more energetic than the second
one, i.e., when either $y < 0.1$ or $y > 0.9$. For a nearly equal
energy sharing ($y \approx$ 0.5), in contrast, accurate relativistic
calculations of the 2M1 and 2E2 rates obviously require summation
over both, the negative and the positive energy states.

Apart from the results of relativistic calculations, we also display
in Fig.~\ref{Fig2} the (sum of the) negative--energy contributions
to the 2M1, 2E2, M1E2 and E2M1 $2s_{1/2} \to 1s_{1/2}$ transition
probabilities as obtained within the semi--relativistic approach
discussed in Section \ref{sub_sect_semi_relativistic}. As expected,
for low--$Z$ ions both the relativistic (dotted line) and
semi--relativistic (dot--dashed line) results basically coincide and
are well described by
Eq.~(\ref{rate_general_nr_negative_M1M1_E2E2_3}). As the nuclear
charge $Z$ is increased, however, semi--relativistic treatment leads
to a slight underestimation of the negative--energy contribution to
the two--photon (differential) transition probabilities. For the
$2s_{1/2} \to 1s_{1/2}$ decay of hydrogen--like uranium ion, for
example, results obtained from
Eq.~(\ref{rate_general_nr_negative_M1M1_E2E2_3}) is about 30 \%
smaller than the corresponding relativistic predictions.

Up to now, we have been considering the $2s_{1/2} \to 1s_{1/2}$
two--photon decay of the hydrogen--like ions. Apart from this
---experimentally well studied---transition, recent theoretical
interest has been focused also on the $2p_{1/2} \to 1s_{1/2}$
\textit{two--photon} decay \cite{LaS05}. Although such a channel is
rather weak comparing to the leading one--photon E1 transition, its
detailed investigation is highly required for future experiments on
the parity violation in simple atomic systems \cite{DrM77}. A number
of calculations \cite{LaS05, LaS06} have been performed, therefore,
for the transition probabilities of the dominant E1M1 and E1E2
multipole components. In order to discuss the role of Dirac's
negative continuum in these calculations, we display in
Fig.~\ref{Fig3} the energy--differential rate for the sum of the
E1M1 and E1E2 $2p_{1/2} \to 1s_{1/2}$ two--photon transitions.
Again, the calculations have been carried out within the Coulomb
gauge for the electron--photon coupling and for three nuclear
charges $Z$ = 1, 54 and 92. As seen from the figure,
negative--energy summation in the second--order transition amplitude
(\ref{S_function_def}) is of great importance for accurate
evaluation of $2p_{1/2} \to 1s_{1/2}$ transition probabilities both
for low--$Z$ and high--$Z$ ions. That is, restriction of the
intermediate--state summation to positive part of Dirac's spectrum
results in an overestimation of the E1M1 and E1E2 differential in
energy decay rates by factors of about 2 and 2.5 for the neutral
hydrogen and hydrogen--like uranium, respectively.

Similarly to the $2s_{1/2} \to 1s_{1/2}$ multipole transitions, we
make use of semi--relativistic formulae from Section
\ref{sub_sect_semi_relativistic} to cross--check our relativistic
computations for the negative--energy contribution to the E1M1 and
E1E2 $2p_{1/2} \to 1s_{1/2}$ decay rates in low--$Z$ domain. Again,
while for neutral hydrogen both, semi--relativistic
(\ref{rate_general_nr_negative_E1M1_E1E2}) and relativistic
approximations produce virtually identical results, they start to
differ as the nuclear charge $Z$ is increased.

So far we have discussed the energy--differential decay rates both
for $2s_{1/2} \to 1s_{1/2}$ and $2p_{1/2} \to 1s_{1/2}$ two--photon
transitions. Integration of these rates over the energy of one of
the photons (see Eq.~(\ref{total_rate})) will yield the
\textit{total} decay rates. In Table I we display the total decay
rates for the various multipole channels of $2s_{1/2} \to 1s_{1/2}$
two--photon decay. In contrast to the photon energy distributions
from above, here relativistic calculations have been performed in
Coulomb (velocity) as well as Babushkin (length) gauges. In both
gauges, negative--energy contribution to the (total) probability of
the leading 2E1 transition is about eight orders of magnitude
smaller than positive--energy term if decay of low--$Z$ ions is
considered but is significantly increased for higher nuclear
charges. For the hydrogen--like uranium, for example, the total 2E1
decay rate is enhanced from 2.9041$× 10^{12}$ s$^{-1}$ in the
velocity gauge and 2.3939$× 10^{12}$ s$^{-1}$ in the length
gauge to the---gauge independent---``exact'' value of 3.8256$×
10^{12}$ s$^{-1}$ if, apart from the positive--energy states, the
Dirac's states with negative energy are taken into account in
transition amplitude Eq.~(\ref{S_function_def}). These results
clearly indicate the importance of the negative--state summation for
the accurate evaluation of 2E1 $2s_{1/2} \to 1s_{1/2}$ total rates
in both, velocity and length gauges. It worth mentioning, however,
that while for velocity gauge our findings are in perfect agreement
with results reported in Ref.~\cite{LaS05} some discrepancy was
found for calculations performed in length gauge for which Labzowsky
and co--workers have argued that the contribution from the Dirac's
negative continuum is negligible even for heaviest ions. The reason
for this discrepancy is not apparent for the moment and, hence,
further investigations are highly required.

In Table I, besides the leading 2E1 decay channel, we present the
results of relativistic calculations for the higher multipole
contributions to the $2s_{1/2} \to 1s_{1/2}$ two--photon transition.
The influence of Dirac's negative continuum is obviously different
for various multipole combinations. While, for example, the
negative--energy contribution to the intermediate--state summation
in low--$Z$ domain is negligible for the E1M2 decay it becomes of
paramount importance for the 2E2 and 2M1 decay channels; an effect
that has been already discussed for the case of the
energy--differential decay rates (see upper panel of
Fig.~\ref{Fig2}). Moreover, $2s_{1/2} \to 1s_{1/2}$ transition with
emission of two magnetic dipole (2M1) photons in light ions seems to
happen almost \textit{exclusively} via the negative energy (virtual)
intermediate states. The total decay rate for this transition
together with the negative--energy contribution to the probability
of the 2E2 channel (evaluated in Coulomb gauge) gives in atomic
units:
\begin{equation}
   w_{2M1} + w^{(-)}_{2E2} = 1.248 \, 10^{-6} \, (\alpha Z)^{10}
   \, ,
\end{equation}
which is in perfect agreement with the semi--relativistic formula
(\ref{total_nr_negative_M1M1_E2E2}).

%
%
%
%
%
\begin{table*}[t]
\begin{center}
\begin{ruledtabular}
\begin{tabular}{|l l|l|l|l|l|l|l|}
 & & \multicolumn{2}{c|}{$Z$=1} & \multicolumn{2}{c|}{$Z$=54} & \multicolumn{2}{c|}{$Z$=92} \\
\cline{3-8}
 & & length & velocity & length & velocity & length & velocity\\
\hline
       & W$_{+}$     &  4.1934 (-05) & 3.2256 (-05)  &
                        3.0381 (+09) & 2.2422 (+09)  &
                        2.1280 (+11) & 1.4126 (+11)\\
E1M1   & W$_{-}$     &  1.9355 (-05) & 9.6773 (-06)  &
                        1.5701 (+09) & 7.7417 (+08)  &
                        1.3745 (+11) & 6.5902 (+10) \\
       & W$_{\rm t}$ &  9.6767 (-06) & 9.6767 (-06)  &
                        6.3731 (+08) & 6.3731 (+08)  &
                        3.8633 (+10) & 3.8633 (+10) \\
\hline
       & W$_{+}$     &  3.6716 (-05) & 1.2339 (-06) &
                        2.6077 (+09) & 8.6699 (+07) &
                        1.7827 (+11) & 6.3367 (+09) \\
E1E2   & W$_{-}$     &  4.5159 (-05) & 9.6769 (-06) &
                        3.1980 (+09) & 6.7698 (+08) &
                        2.1653 (+11) & 4.4598 (+10) \\
       & W$_{\rm t}$ &  6.6117 (-06) & 6.6117 (-06) &
                        4.2942 (+08) & 4.2942 (+08) &
                        2.3584 (+10) & 2.3584 (+10) \\
\end{tabular}
\end{ruledtabular}
\label{Table1} \caption{Total rates (in s$^{-1}$) for the several
multipole combinations of $2p_{1/2} \to 1s_{1/2}$ two--photon
decay.}
\end{center}
\vspace{-0.6cm}
\end{table*}

As mentioned above for the computation of the photon energy
distributions in low--$Z$ domain, negative--energy contribution to
the intermediate--state summation is rather pronounced not only for
the higher multipole terms of $2s_{1/2} \to 1s_{1/2}$ decay but also
for the leading E1M1 and E1M2 (two--photon) channels of $2p_{1/2}
\to 1s_{1/2}$ transition. Our relativistic calculations displayed in
Table II indicate that one should account for negative--continuum
summation also for an accurate evaluation of the \textit{total}
decay rates for these two decay channels. For the decay of light
elements, sizable contribution from the negative--continuum
intermediate states arises both in length and velocity gauges.
Again, these results partially question the predictions by Labzowsky
and co--workers \cite{LaS05} who claimed a minor role of negative
energy terms for E1M1 and E1M2 calculations in length gauge. For the
velocity gauge, in contrast, our relativistic calculations:
\begin{equation}
  w^{(-)}_{E1M1} + w^{(-)}_{E1E2} = 5.822 \, 10^{-5} \, (\alpha Z)^{8}
   \, ,
\end{equation}
are in a good agreement both, with the semi--relativistic prediction
(\ref{total_nr_negative_E1M1_E1E2}) and data presented in
Ref.~\cite{LaS05}.

%
%
%
%
\section{Summary and outlook}
\label{Sec_summary}

In conclusion, the two--photon decay of hydrogen--like ions has been
re--investigated within the framework of second--order perturbation
theory, based on Dirac's relativistic equation. Special attention
has been paid to the summation over the intermediate ionic states
which occurs in such a framework and runs over \textit{complete}
one--particle spectrum, including a summation over discrete (bound)
states as well as the integration over the positive and negative
continua. In particular, we discussed the role of the
\textit{negative} energy continuum in an accurate evaluation of the
second--order transition amplitudes and, hence, the
energy--differential as well as total decay rates. Detailed
calculations of these rates have been presented for the $2s_{1/2}
\to 1s_{1/2}$ and $2p_{1/2} \to 1s_{1/2}$ two--photon transitions in
neutral hydrogen as well as hydrogen--like xenon and uranium ions.
As seen from the results obtained, both the total decay
probabilities and the energy distributions of the simultaneously
emitted photons can be strongly affected by the negative--state
summation not only for heavy ions but also for low--$Z$ domain. We
demonstrate, however, that the role of Dirac's negative continuum
becomes most pronounced for the higher (non--dipole) terms in the
expansion of the electron--photon interaction; similar effect has
been recently reported for the theoretical description of
hydrogen--like systems exposed to intense electromagnetic pulses
\cite{SeL09}.

In the present work, we have restricted our discussion of the
negative energy contribution to the second--order calculations of
the total and energy--differential decay rates. Even stronger
effects due to the Dirac's negative continuum can be expected,
however, for the angular and polarization correlations between
emitted photons. Theoretical investigation of these correlations
which requires also detailed analysis of interference terms between
the various (two--photon) multipole combinations is currently
underway and will be published soon.

\section*{Acknowledgements}

A.S. acknowledges support from the Helmholtz Gemeinschaft and GSI
under the project VH--NG--421.  This research was supported in part
by FCT Project No.  POCTI/0303/2003 (Portugal), financed by the
European Community Fund FEDER, by the French-Portuguese
collaboration (PESSOA Program, Contract No. 441.00), and by the
Acções Integradas Luso-Francesas (Contract No. F-11/09) and
Luso-Alemãs (Contract No. A-19/09). Laboratoire Kastler Brossel is `` Unité Mixte de Recherche du CNRS, de l' ENS et de l' UPMC No.
8552.'' P.A. acknowledges the support of the FCT, under Contract No.
SFRH/BD/37404/2007. This work was also partially supported by
Helmholtz Alliance HA216/EMMI.

%
%
%
%

\end{document}